\documentclass[12pt]{amsart}

\usepackage[utf8]{inputenc}
\usepackage[T1]{fontenc}

\usepackage[dvipsnames]{xcolor}
\usepackage{amsmath, amsfonts, amsthm, amssymb}
\usepackage[top=1in, bottom=1in, left=1in, right=1in]{geometry}
\usepackage[
backref=page, colorlinks=true, allcolors=blue]{hyperref}
\usepackage[alphabetic,abbrev,lite,backrefs]{amsrefs} 
\usepackage{tikz}
\usepackage{changepage}
\usepackage{mathrsfs}
\usepackage{tikz-cd}
\usepackage{mathtools} 
\usepackage{bm} 
\usepackage{caption}
\usepackage{blkarray} 
\usepackage{enumitem} 
\usepackage{algorithm2e}
\usepackage{multirow}

\usepackage[colorinlistoftodos,prependcaption,textsize=scriptsize]{todonotes}
\newcommand{\indep}{\perp \!\!\! \perp}

\makeatletter
\providecommand\@dotsep{5}
\renewcommand{\listoftodos}[1][\@todonotes@todolistname]{%
  \@starttoc{tdo}{#1}}
\makeatother

\theoremstyle{plain}
	\newtheorem{theorem}{Theorem}[section]
    \newtheorem*{theorem*}{Theorem}

    \newtheorem*{Question*}{Question}
\theoremstyle{definition}
    
    \newtheorem*{defn*}{Definition}
    \newtheorem{example}[theorem]{Example}
    \newtheorem*{example*}{Example}
    
\theoremstyle{remark}
	\newtheorem{remark}[theorem]{Remark}
	\newtheorem*{remark*}{Remark}

\AtBeginEnvironment{example}{%
  \pushQED{\qed}%
}
\AtEndEnvironment{example}{\popQED\endexample}

\usepackage{mathtools}

\DeclarePairedDelimiterX\Set[1]\{\}{#1}

\newcommand{\mc}[1]{\mathcal{#1}}

\def\P{{\mathbb{P}}}

\def\M{{\mc{M}}}

\DeclareMathOperator{\dlog}{dlog}

\title{LikelihoodGeometry: Macaulay2 Package}

\author{David Barnhill}
\address{Department of Mathematics, United States Naval Academy, Annapolis, MD}
\email{barnhill@usna.edu}

\author{John Cobb}
\address{Department of Mathematics, Auburn University, Auburn, AL}
\email{jdcobb3@gmail.com}

\author{Matthew Faust}
\address{Department of Mathematics, Michigan State University, East Lansing, MI}
\email{mfaust@msu.edu}

\begin{document}

\maketitle
\vspace{-0.3in}
\begin{abstract}
    This note introduces the \texttt{LikelihoodGeometry} package for the computer algebra system \textit{Macaulay2}. This package gives tools to construct the likelihood correspondence of a discrete algebraic statistical model, a variety that that ties together data and their maximum likelihood estimators. This includes methods for constructing and combining popular statistical models and calculating their ML-degree.
\end{abstract}

\section{Introduction and Background}
Maximum likelihood estimation (MLE) is a fundamental computational problem in statistics with recent insights stemming from studying the geometry of discrete algebraic statistical models as varieties \cites{Catanese2004TheML, huh2013maximum, LikelihoodGeometry,Amendola2019MaximumLE,Maxim2022LogarithmicCB}. The likelihood correspondence of a discrete algebraic statistical model $X$ is the universal family of solutions to all MLE problems on $X$. This variety, or equivalently its vanishing ideal the \textit{likelihood ideal}, encodes all information about the MLE problem \cite[Section 2]{Hosten2004SolvingTL}, such as the {\em maximum likelihood} (ML) degree of the model. In \cite{barnhill2023likelihood}, we presented algorithms to construct the likelihood ideal of toric models and found a Gr\"{o}bner basis in the case of complete and joint independence models arising from multi-way contingency tables. The \textit{Macaulay2} \cite{M2} package \texttt{LikelihoodGeometry} implements these new algorithms alongside functions which allow the construction of various sorts of models and a calculation of their ML-degree. A more general algorithm \cite[Algorithm 6]{Hosten2004SolvingTL} for non-toric models is also implemented. This package differs from \texttt{GraphicalModelsMLE} \cite{amendola2022computing} by constructing MLE estimates for \textit{discrete} algebraic statistical models (rather than Gaussian) by directly constructing the likelihood ideal rather than numerically solving the problem for specified data vectors.

For the rest of this section, we give some background for the models and likelihood correspondence and introduce the main functions of the package with examples. In Section \ref{sec: Constructing the LC}, we detail the implementation of the main algorithm $\texttt{computeLC}$.

\subsection{Likelihood Geometry}\label{sec: Background on Likelihood Geometry}
Fix a system of homogeneous coordinates $p_0, p_1, \dots, p_n$ of complex projective space $\P_p^n$ representing the space of probabilities. Similarly, we consider $u$ as coordinates of the data space $\P^n_u$. The likelihood function on $\P_p^n$ is given by
\begin{equation*}
    \ell_u(p) = \frac{p_0^{u_0}p_1^{u_1}\cdots p_n^{u_n}}{(p_0+p_1+\cdots + p_n)^{u_0+u_1+\cdots+u_n}}.
\end{equation*}
We aim to study the MLE problem on the restriction of $\ell_u$ to a given closed irreducible subvariety $\M\subseteq \P_p^n$. Fixing a data vector $u$, maximum likelihood estimates are given by the maxima of $\ell_u$. These maxima belong to the critical set of the log--likelihood function $\log \ell_u$. This critical set is given by $V(\dlog \ell_u) \subset \P^n_p$, the vanishing set of the gradient of $\log \ell_u$. By varying $u$, we can collect such critical sets into a universal family $\mathcal{L}_\M \subset \P_p^n \times \P_u^n$ called the \textit{likelihood correspondence}. More information can be found in \cite[Section 2]{barnhill2023likelihood} or in \cite[Section 2]{Hosten2004SolvingTL}.

The main function of the package is \texttt{computeLC}, which can take in an arbitrary ideal representing the vanishing ideal of a statistical model and returns the likelihood ideal. Depending on the input, it automatically implements the fastest algorithms from Section \ref{sec: Constructing the LC}.

\begin{example}\label{ex:HardyWeinberg}
    Consider the problem of flipping a biased coin two times and recording the number of heads. Let $p=(p_0,p_1,p_2)$ represent the vector of probabilities of getting $0$ heads, $1$ head, and $2$ heads respectively. The possible probabilities for this experiment are defined by the \textit{Hardy--Weinberg curve} $\M = V(4p_0p_2-p_1^2) \subset \P_p^2$ \cite{Alg_stat_CB}, a statistcal model studied in population genetics. The likelihood correspondence is a surface $\mathcal{L}_\M \subset \P^2_p \times \P^2_u$ with ML-degree 1: 
    
    \begin{verbatim}
    i1 : R = CC[p_0,p_1,p_2];
    i2 : M = ideal(4*p_0*p_2-p_1^2);
    o2 : Ideal of R
    i3 : L = computeLC(M)  
    o3 = ideal (4p u  - p u  + 2p u  - 2p u , 2p u  - 2p u  + p u  - 4p u ,
                  2 0    1 1     2 1     1 2    1 0     0 1    1 1     0 2
                 2
                p  - 4p p )
                 1     0 2 
    o3 : Ideal of CC[p ..p , u ..u ]
                      0   2   0   2
    i4 : MLdegree(M)
    o4 = 1 
    \end{verbatim} \vspace{-0.5in} \qedhere
\end{example}

\subsubsection{Discrete Random Variables} 
All of the analyses in this paper applies to \textit{discrete} algebraic statistical models. This package implements a new data type \texttt{DiscreteRandomVariable} which holds an \texttt{arity} and can be given a custom probability mass function as \texttt{pmf} (default is uniform). There are methods to \texttt{sample}, take the \texttt{mean}, and see the state space with \texttt{states}. 

\begin{example} The following code block demonstrates some of the methods available for \texttt{DiscreteRandomVariable}.
    \begin{verbatim}
    i1 : g = new HashTable from {
         1 => 0.5,
         2 => 0.3,
         3 => 0.2
     };
    i2 : f = x -> if g#?x then g#x else 0;
    i3 : Y = discreteRandomVariable(3, f);
    o3 = Y
    o3 : DiscreteRandomVariable
    i4 : {states Y, mean Y, sample(Y,2)}
    o4 = {{1, 2, 3}, 1.7, {1, 1}}
    o4 : List
    \end{verbatim} \vspace{-0.5in} \qedhere
\end{example}

\subsection{Different Types of Models}
\subsubsection{Toric Models} A toric model is simply a model which is a toric variety and are among the most thoroughly studied statistical models \cites{on_tor, Alg_stat_CB, sullivant}. One method to construct a toric model is to pass in a matrix $A$ such that each column is the lattice vertex of the corresponding polytope to the function \texttt{toricModel}, which returns a \texttt{NormalToricVariety}. The function \texttt{toricIdeal} can take in a toric model given by a \texttt{NormalToricVariety} or the defining matrix and returns the vanishing ideal.

\begin{example}
    The rational normal curve of degree $d$ is the image of the map
    \begin{align*}
        \varphi: &\hspace{0.15in} \P^1 \longrightarrow \P^d\\
        &[s:t] \mapsto [s^d: s^{d-1}t : \cdots: st^{d-1}: t^d]
    \end{align*}
    and is defined as a toric variety by the $2\times (d+1)$ matrix 
    \[ \begin{bmatrix}
        1 & 1 & 1 & \cdots & 1 \\
        1 & 2 & 3 & \cdots & d+1
    \end{bmatrix}.\]
    Taking $d=3$ gives the twisted cubic:
    \begin{verbatim}
    i1 : A = matrix{{1,1,1,1},{1,2,3,4}};
                   2       4
    o1 : Matrix ZZ  <-- ZZ
    i2 : X = toricModel(A);
    i3 : R = QQ[p_0,p_1,p_2,p_3];
    i4 : M = minors(2, matrix{{p_0,p_1,p_2},{p_1,p_2,p_3}}); --twisted cubic
    o4 : Ideal of R
    i5 : M == toricIdeal(X,R)
    o5 = true
    \end{verbatim} \vspace{-0.5in} \qedhere
\end{example}

Rational normal curves are part of the larger class of rational normal scrolls, and we have implemented the construction of this class via \texttt{rationalNormalScroll}. The family of rational normal scrolls can be parametrized by a list of integers $a_0, \dots, a_k$ corresponding to $S(a_0,\dots, a_k) \subseteq \P^N$ where $N=\sum_{i=0}^k (a_i) + k$ \cite{petrovic08}. For example, the rational normal curve of degree $d$ can be constructed using \texttt{rationalNormalCurve(\{d+1\})}. Without a toric polytope on hand, one can construct it from a toric ideal using \texttt{toricPolytope}. 

\begin{verbatim}
    i6 : toricPolytope(M)
    o6 = | 1  1  1 1 |
         | -2 -1 0 1 |
                  2       4
    o6 : Matrix ZZ  <-- ZZ
\end{verbatim}

Using this package, one can construct the likelihood ideal of toric models much faster than arbitrary ideals using the algorithm detailed in Section \ref{sec: Constructing the LC}.

\begin{verbatim}
    i1 : X = rationalNormalScroll({2,2,3});
    i2 : elapsedTime computeLC(X);
     -- 0.0591624 seconds elapsed
    o3 : Ideal of QQ[p ..p , u ..u ]
                      0   6   0   6
    i4 : elapsedTime computeLC(toricIdeal(X)); -- > 3600 seconds elapsed
\end{verbatim}

\begin{verbatim}
    
\end{verbatim}

\subsubsection{Hierarchical Log-Linear Models}
Within the class of toric models is the class of hierarchical log-linear models. Let $\mc{X} = [d_1] \times \cdots \times [d_n]$ denote the joint state space of the discrete random variables $X_1,\dots, X_n$. A \textit{hierarchical log--linear model} is defined by a collection $\{G_1,\dots, G_g\}$ of non-empty subsets of $X = \{X_1,\dots, X_n\}$ called \textit{generators}.  

\begin{example}\label{ex: 3chain}
    Let $X = \{a,b,c\}$ be a list of binary random variables and the generators be $\{a,b\}$ and $\{b,c\}$. The function \texttt{makeLogLinearMatrix} constructs the defining matrix for the corresponding toric model.
    \begin{verbatim}
    i1 : a = discreteRandomVariable 2; b = discreteRandomVariable 2; 
         c = discreteRandomVariable 2;
    i4 : X = {a,b,c};
    i5 : G = {{a,b}, {b,c}};
    i6 : makeLogLinearMatrix(G,X)
    o6 = | 1 1 0 0 0 0 0 0 |
         | 0 0 1 1 0 0 0 0 |
         | 0 0 0 0 1 1 0 0 |
         | 0 0 0 0 0 0 1 1 |
         | 1 0 0 0 1 0 0 0 |
         | 0 1 0 0 0 1 0 0 |
         | 0 0 1 0 0 0 1 0 |
         | 0 0 0 1 0 0 0 1 |
                  8       8
    o6 : Matrix ZZ  <-- ZZ
    \end{verbatim}
    This defines the toric model representing the conditional independence statement $a \indep c \, | \, b$, or ``$a$ is independent of $c$ given $b$''.
\end{example}

\subsubsection{Undirected Graphical Models} 
Within the class of hierarchical log-linear models are the class of undirected graphical models. The undirected graphical model for a graph $\mc{G}$ is the log--linear model on $X$ in which the generators are cliques (maximal complete subgraphs) of $\mc{G}$. Many functions described so far can take a graph whose vertices are discrete random variables as input, including \texttt{toricModel}, \texttt{toricIdeal}, and \texttt{makeLogLinearMatrix}.

\begin{example}
    The model in Example \ref{ex: 3chain} is the undirected graphical model corresponding to the following graph:
    
    
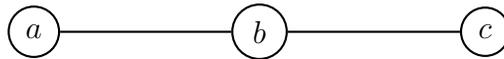
\begin{figure}[h]
        \centering
    \begin{tikzpicture}[node distance={30mm}, thick, main/.style = {draw, circle}]
    \node[main] (3) {$a$}; 
    \node[main] (4) [right of=3] {$b$};
    \node[main] (5) [right of=4] {$c$}; 
    
    \draw [-](3) -- (4);
    \draw [-](4) -- (5);
    \end{tikzpicture}
    \end{figure}
    The toric model corresponding to a graph can be computed using \texttt{toricModel}, and the ideal using \texttt{toricIdeal}.
    \begin{verbatim}
    i1 : a = discreteRandomVariable 2; b = discreteRandomVariable 2; 
          c = discreteRandomVariable 2;
    i4 : G = graph{{a,b}, {b,c}};
    i5 : X = toricModel G
    o5 = X
    o5 : NormalToricVariety
    i6 : I = toricIdeal G
    o6 = ideal (p p  - p p , p p  - p p )
                  3 6    2 7   1 4    0 5
    o6 : Ideal of QQ[p ..p ]
                       0   7
    \end{verbatim}\vspace{-0.5in} \qedhere
\end{example} 

\section{Constructing the Likelihood Correspondence}\label{sec: Constructing the LC}
Upon running \texttt{computeLC}, there are three possible algorithms that will run. In the case that the input is an ideal, \cite[Algorithm 6]{Hosten2004SolvingTL} utilizing Lagrange multipliers has been implemented. In the case that a toric model is used, then the faster Algorithm \ref{alg: toric} is used. Finally, if a undirected graphical model is input, then a check is performed to see if it is a complete or joint independence model in which case we return a Gröbner basis as given in Theorem 3.2 and Corollary 3.5 in \cite{barnhill2023likelihood}.

\begin{algorithm}[h]
    \caption{Constructing likelihood ideals of toric models}\label{alg: toric}  
    \SetKwInOut{Input}{Input}\SetKwInOut{Output}{Output}
    \Input{a \texttt{NormalToricVariety} $X$ representing a toric model}
    \Output{the likelihood ideal}
    \eIf{$X$ {\normalfont is a graphical model whose components are complete graphs}}{ \Return Gröbner basis as given in \cite{barnhill2023likelihood}\;}{
    $A \coloneqq$ defining matrix of $X$\; 
    $R \coloneqq$ \texttt{LCRing}($X$)\;
    $I_X \coloneqq$ \texttt{toricIdeal}($A$, $R$)\;
    $M \coloneqq$ \texttt{reshape}($R^\texttt{numcol}$, $R^2$, \texttt{matrix}({\texttt{gens} $R$}))\;
    \Return \texttt{saturate}($I_X + $\texttt{minors}(2, $A*M$), $(\sum p_i)(\prod p_i))$
    }
\end{algorithm}

\texttt{MLdegree} is then implemented by taking the fiber of a generic map from the likelihood correspondence down to the model.

\begin{remark}
    As noted in \cite{barnhill2023likelihood}, it appears to suffice to saturate by only the single hyperplane $V(\sum p_i)$ in the final step, which can be significantly faster. This is how it is implemented and correctness is guaranteed if the output is prime.
\end{remark}

\subsection*{Data Availability} All code is available as the \textit{Macaulay2} package, \texttt{LikelihoodGeometry}.\footnote{ \href{https://github.com/johndcobb/LikelihoodGeometry}{https://github.com/johndcobb/LikelihoodGeometry}}

\subsection*{Acknowledgement}
Part of this research was performed while the authors were visiting the Institute for Mathematical and Statistical Innovation (IMSI), which is supported by the National Science Foundation (Grant No. DMS-1929348). J. Cobb is supported by DMS-2402199. M. Faust is supported by DMS-2052519. The implementation of the functions \texttt{toricIdeal} and \texttt{toricPolytope} are from the packages \texttt{SlackIdeals} \cite{Macchia2020SlackII} and \texttt{Quasidegrees} \cite{Barrera2015ComputingQO}, respectively. 
\bibliography{refs}{}

\end{document}